\documentclass[twocolumn,showpacs,preprintnumbers,amsmath,amssymb,prl,aps,superscriptaddress,reprint]{revtex4-1}

\usepackage{graphicx}% Include figure files
\usepackage{dcolumn}% Align table columns on decimal point
\usepackage{bm}% bold math
\usepackage{color}

\newcommand{\ba}{BaFe$_2$As$_2$}

\newcommand{\caco}{Ca(Fe$_{1-x}$Co$_x$)$_2$As$_2$}
\newcommand{\bat}{Ba(Fe,TM)$_2$As$_2$}

\newcommand{\bani}{Ba(Fe$_{1-x}$Ni$_x$)$_2$As$_2$}
\newcommand{\bacu}{Ba(Fe$_{1-x}$Cu$_x$)$_2$As$_2$}
\newcommand{\nfe}{$\langle n_{Fe}\rangle$}

\begin{document}

\title{Observation of charge accumulation and onsite Coulomb repulsion at transition metal impurities in the iron pnictides}
\author{R. Kraus}
\author{V. Bisogni}
\author{L. Harnagea}
\author{S. Aswartham}
\author{S. Wurmehl}
\affiliation{Leibniz Institute for Solid State and Materials Research, Helmholtzstrasse 20, D-01171 Dresden, Germany}

\author{G. Levy}
\affiliation{Department of Physics and Astronomy, University of British Columbia, 6224 Agricultural Road, British Columbia V6T 1Z1, Vancouver, Canada}

\author{I. S. Elfimov}
\affiliation{Department of Physics and Astronomy, University of British Columbia, 6224 Agricultural Road, British Columbia V6T 1Z1, Vancouver, Canada}

\author{B. B\"uchner}
\affiliation{Leibniz Institute for Solid State and Materials Research, Helmholtzstrasse 20, D-01171 Dresden, Germany}
\affiliation{Institute for Solid State Physics, Dresden Technical University, TU-Dresden, 01062 Dresden, Germany} 
\author{G.A. Sawatzky}
\affiliation{Department of Physics and Astronomy, University of British Columbia, 6224 Agricultural Road, British Columbia V6T 1Z1, Vancouver, Canada}

\author{J. Geck}
\email{j.geck@ifw-dresden.de}
\affiliation{Leibniz Institute for Solid State and Materials Research, Helmholtzstrasse 20, D-01171 Dresden, Germany}

\begin{abstract}
We report a combined valence band photoemission and Auger spectroscopy study of single crystalline Ca(Fe,Co)$_2$As$_2$  and  Ba(Fe,TM)$_2$As$_2$ with TM=Ni or Cu.  The valence band photoemission data show directly that the TM-states move to higher binding energies with increasing atomic number, contributing less and less to the states close to the Fermi level. Furthermore, the  3d$^8$ final state of the $LVV$ Auger decay, which is observed for Ni and Cu, unambiguously reveals the accumulation of charge at these impurities. We also show that the onsite Coulomb interaction on the impurity strongly increases when moving from Co over Ni to Cu. Our results quantify the impurity potentials 
%of Co, Ni and Cu 
and imply that the superconducting state is robust against impurity scattering.
\end{abstract}

\date{\today}
\maketitle

High-temperature superconductivity (HTS) in the iron pnictides is one of the most intensively studied topics in current condensed matter science. Although research efforts established many properties of these materials, important questions still remain to be clarified\,\cite{Stewart2011}. One of these questions regards the effect of replacing a few percent of Fe by other transition metals (TMs) like Co, Ni or Cu. The substitution of Fe by Co or Ni induces HTS\,\cite{Sefat2008} and therefore it is  very important to understand the effects of the TM-impurities on the electronic structure. Exactly these effects, however, remain controversial  (e.g. Refs.\,\onlinecite{Berlijn2012},\onlinecite{McLeod2012} and references therein).

It is often assumed that Co, Ni and Cu mainly dope electrons into the Fe bands without seriously affecting the electronic structure otherwise\,\cite{Canfield2010}. But this interpretation has been challenged recently. While some studies confirm that the TM substitutions indeed dope additional charge carriers into the Fe-bands\,\cite{Liu2011a,Malaeb2009,Konbu2011}, others report the lack of additional charge at the Fe sites\,\cite{Bittar2011,Merz2012,Khasanov2011}, casting doubts on the electron doping via TM substitution. 
%\cite{Sefat2008}
In addition, recent theoretical studies show that the random scattering potential introduced by the substituted TMs has important effects, which are often neglected but may in fact cause significant modifications to the electronic system\,\cite{Berlijn2012,Haverkort2011,Sefat2008,Wadati2010}. Hence, there is a considerable ongoing debate about the effects of TM substitutions in the iron pnictides.

In this letter we shed light on this issue and investigate the effects of Co,  Ni and Cu-substitution %on the electronic structure of the iron pnictides
%\ca\ and  \ba\ 
by means of photoemission spectroscopy (PES), Auger electron spectroscopy (AES) as well as model calculations. 
%This allows to quantify the scattering potentials caused by the substituted TMs and to directly observe the accumulation of charge at these impurities.
%
Specifically we report studies of Ca(Fe,Co)$_2$As$_2$ and \bat\ (TM=Ni, Cu). For Co we chose the Ca- instead of  Ba-material, in order to avoid complications due to the overlap of the Ba $M_{4,5}$ and the Co $L_{2,3}$ edges\,\cite{McLeod2012}. 
%
%For all three systems we determine the Coulomb repulsion on the TM impurities and observe directly the average binding energy as well as the accumulation of charge for Ni and Cu. 
%
The experimental data were collected at the UE52-PGM beamline of the HZB-BESSY II synchrotron source. Single crystalline samples were grown as described elsewhere\,\cite{Harnagea2011,Aswartham2011a} and cleaved {\it in situ} just before the measurements, which resulted in fresh and mirror-like surfaces. 
More details about the sample characterization can be found in the supplementary information.
PES and AES spectra were recorded using a  Scienta R4000 hemispherical analyzer, whereas the X-ray absorption (XAS)  was measured via the total electron and fluorescence yield. The photon energy was calibrated against the Au $4f$ lines, which means that the measured binding energies always refer to the chemical potential. All spectra where recorded with the photon beam at an incident angle of 35$^\circ$ with respect to the sample surface, which was parallel to the FeAs-layers.
For the PES and AES measurements horizontal and vertical polarized light was used, respectively, to optimize the relative intensities of the two components\,\cite{Weinelt1997}.

\begin{figure}
\centering
%\vspace{15cm}
\includegraphics[width=0.725\columnwidth]{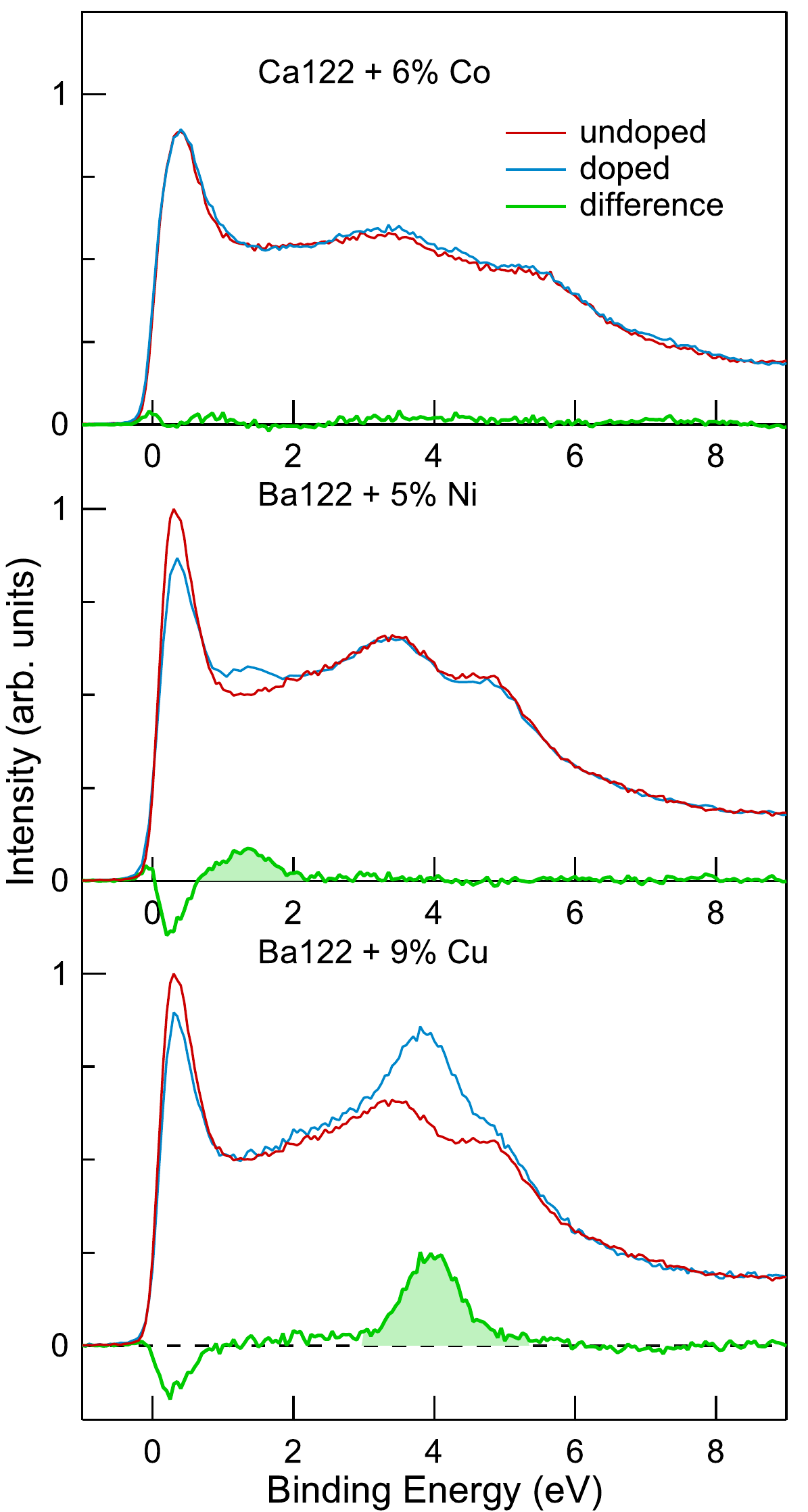}
\caption{Valence band photoemission data for \ba, CaFe$_2$As$_2$, \caco\ with $x=0.06$, \bani\ with $x=0.05$, and \bacu\ with $x=0.09$. All measurements were performed at room temperature using a photon energy of $h \nu=600$\,eV and normalized  by the area above of 7\,eV binding energy.
}\label{fig:PES}
\end{figure}

The valence band photoemission data for  Ca(Fe,Co)$_2$As$_2$ and \bat\ (TM=Ni, Cu)  are presented in Fig.\,\ref{fig:PES}, including the spectra of the corresponding pure parent compounds. In the case of the parent materials, the PES intensity at 0-2\,eV and 3-6\,eV is due to mostly Fe:$3d$ and As:$4p$ derived bands\,\cite{Jong2009,Koitzsch2010}. The comparison of \caco\ with $x=0.06$ to its parent compound CaFe$_2$As$_2$, reveals that both spectra are identical within the error of the experiment.
However, for the Ni- and Cu-substituted materials, additional features can be  observed clearly in the valence band photoemission. An additional structure centered at the binding energy $\langle\epsilon_B^{Ni}\rangle=1.4$\,eV is present in the Ni-substituted \ba , which does not exist in the parent material and which is evident in the difference of the two spectra. Even more striking, a strong additional peak at $\langle\epsilon_B^{Cu}\rangle=4$\,eV is found for the Cu-substituted material. We stress that these additional features were fully reproducible and observed for different samples.

\begin{figure*}
\centering
\includegraphics[width=0.32\textwidth]{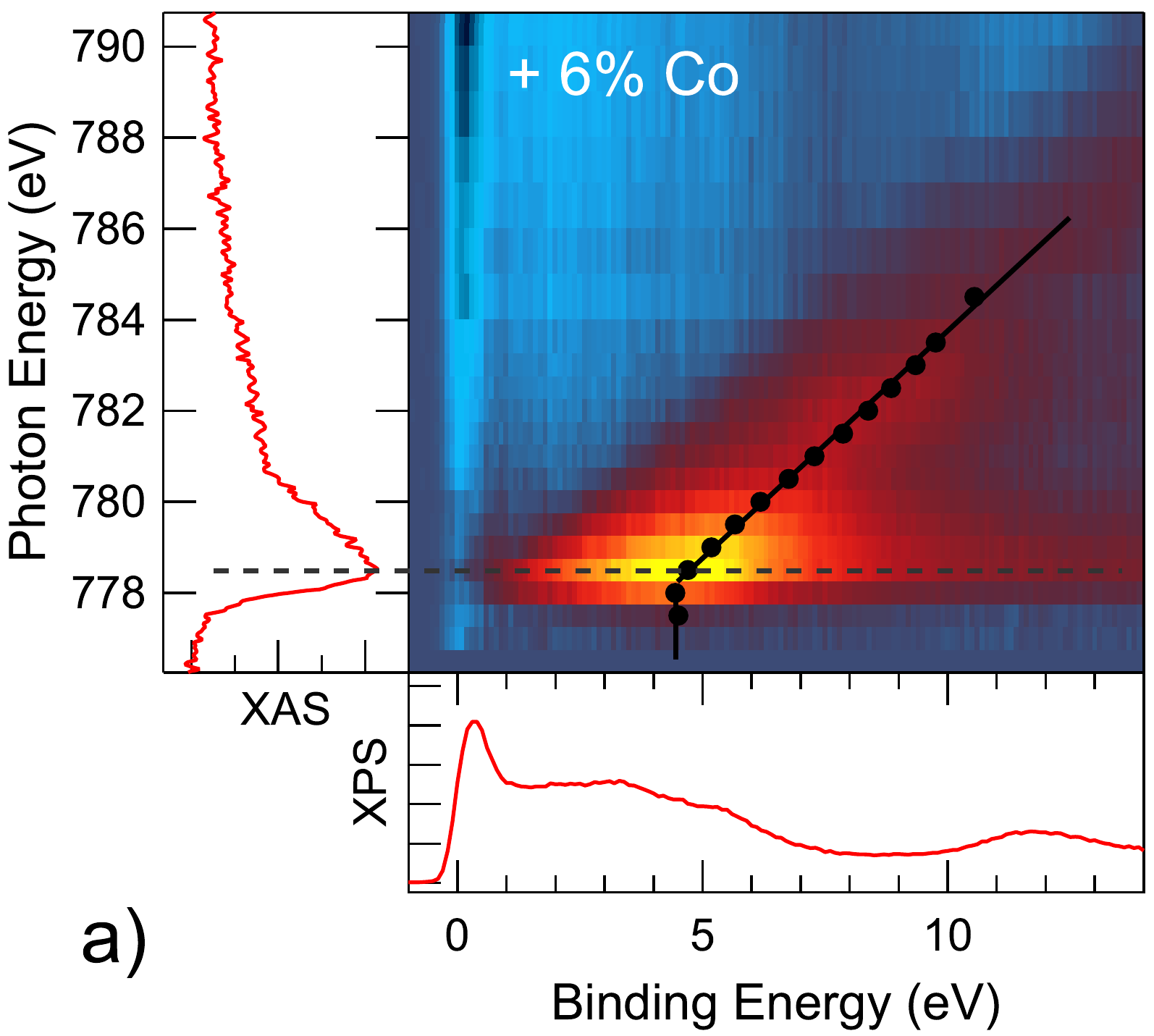}
\includegraphics[width=0.32\textwidth]{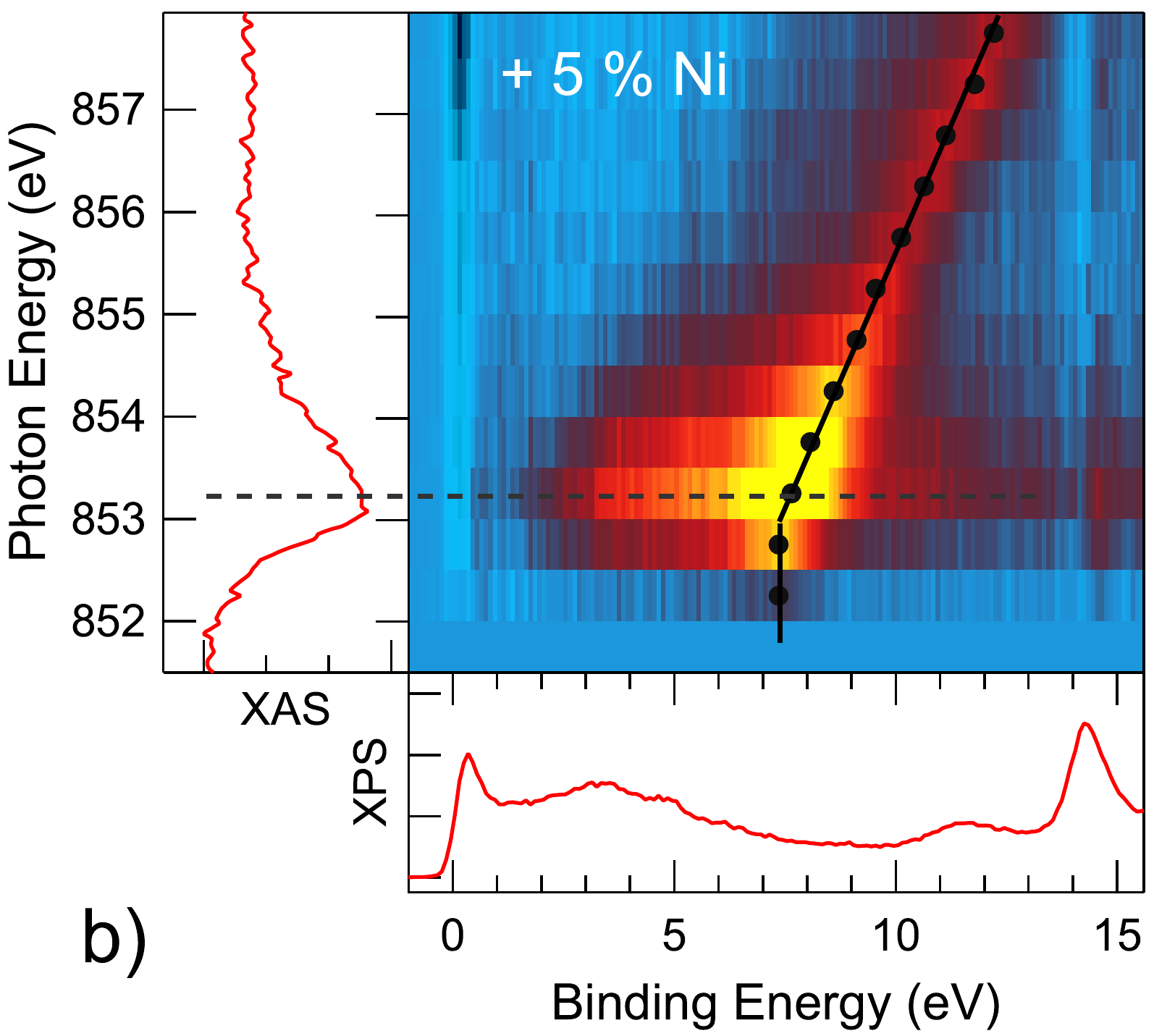}
\includegraphics[width=0.32\textwidth]{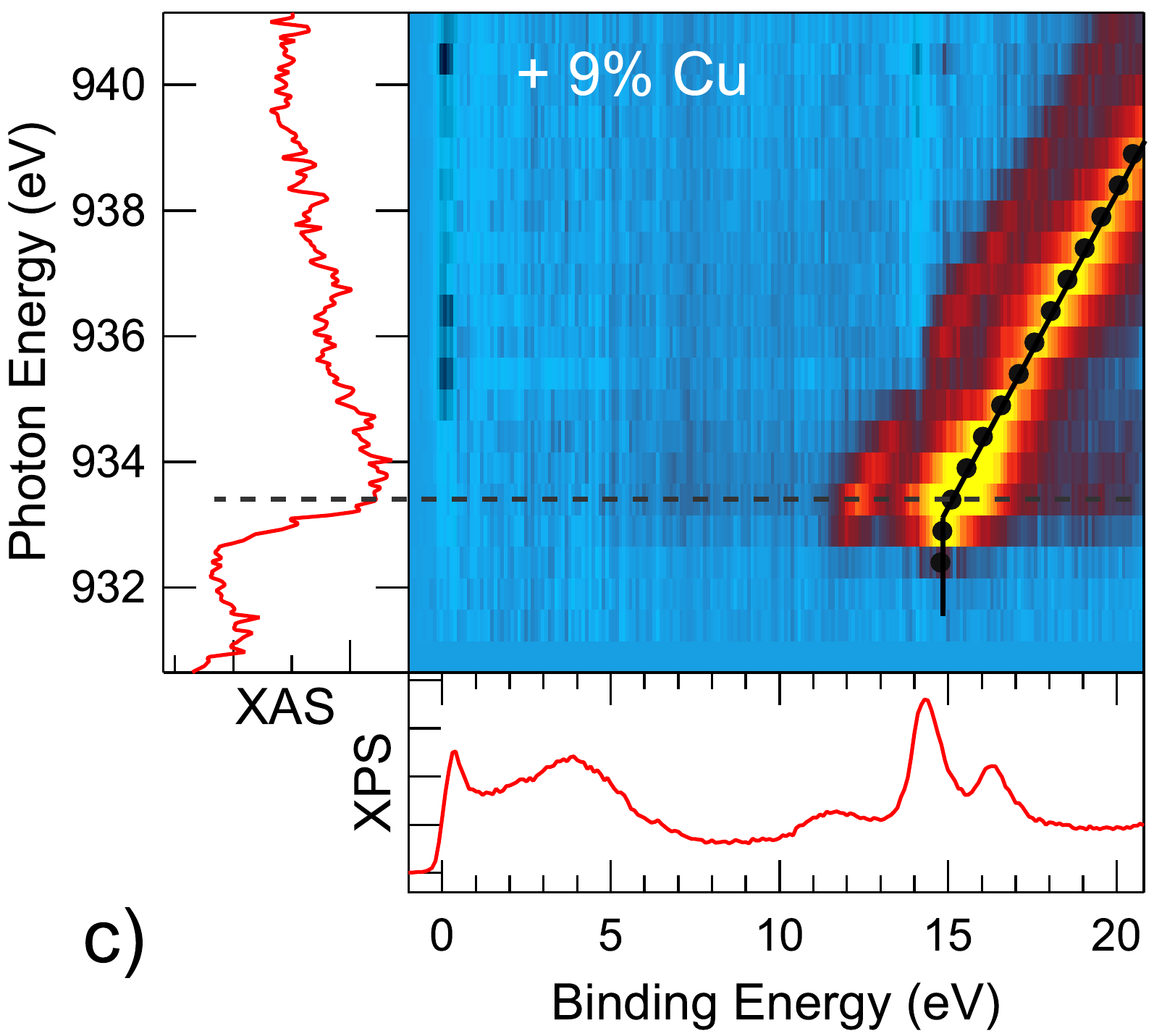}

\includegraphics[width=0.32\textwidth]{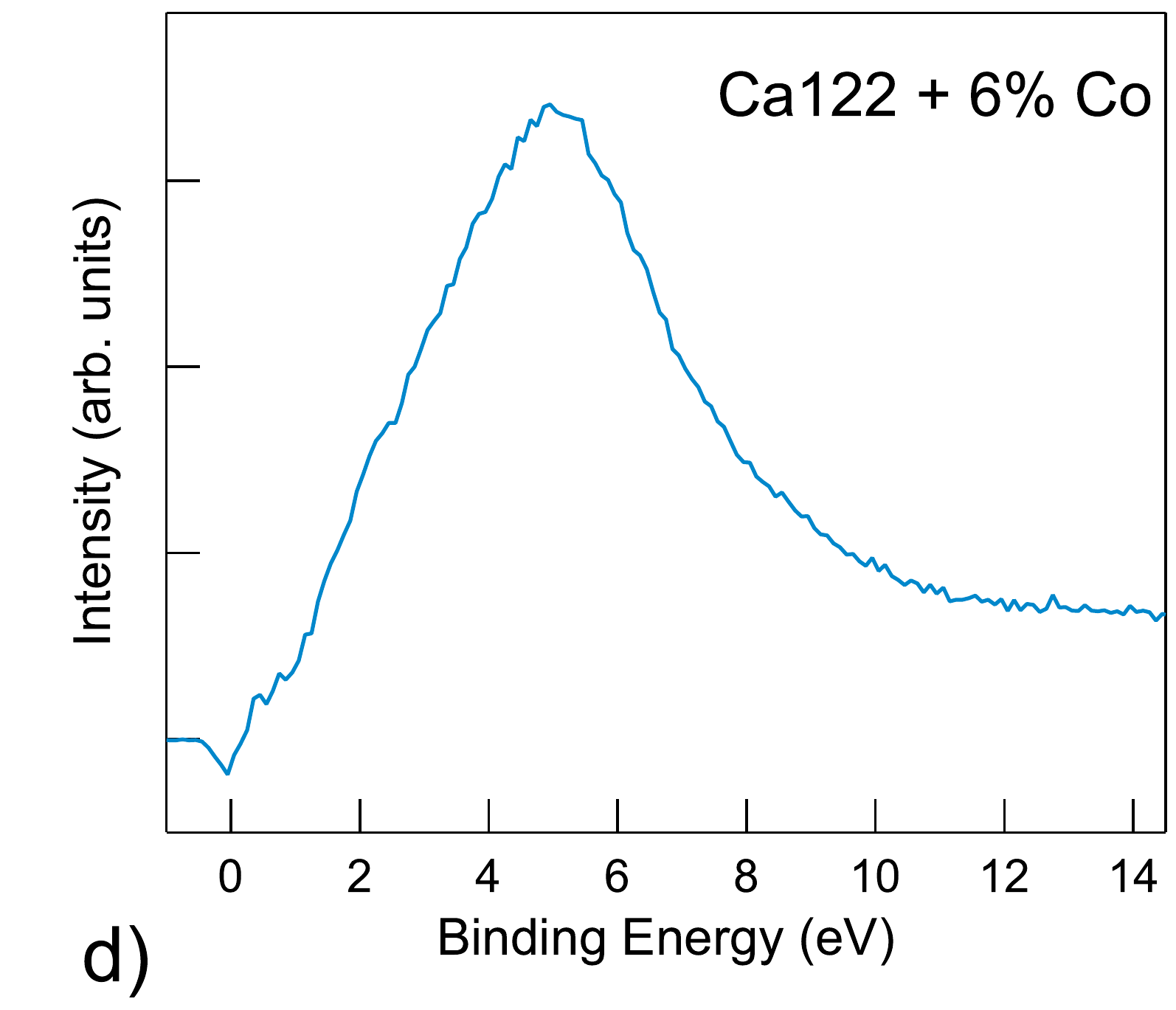}
\includegraphics[width=0.32\textwidth]{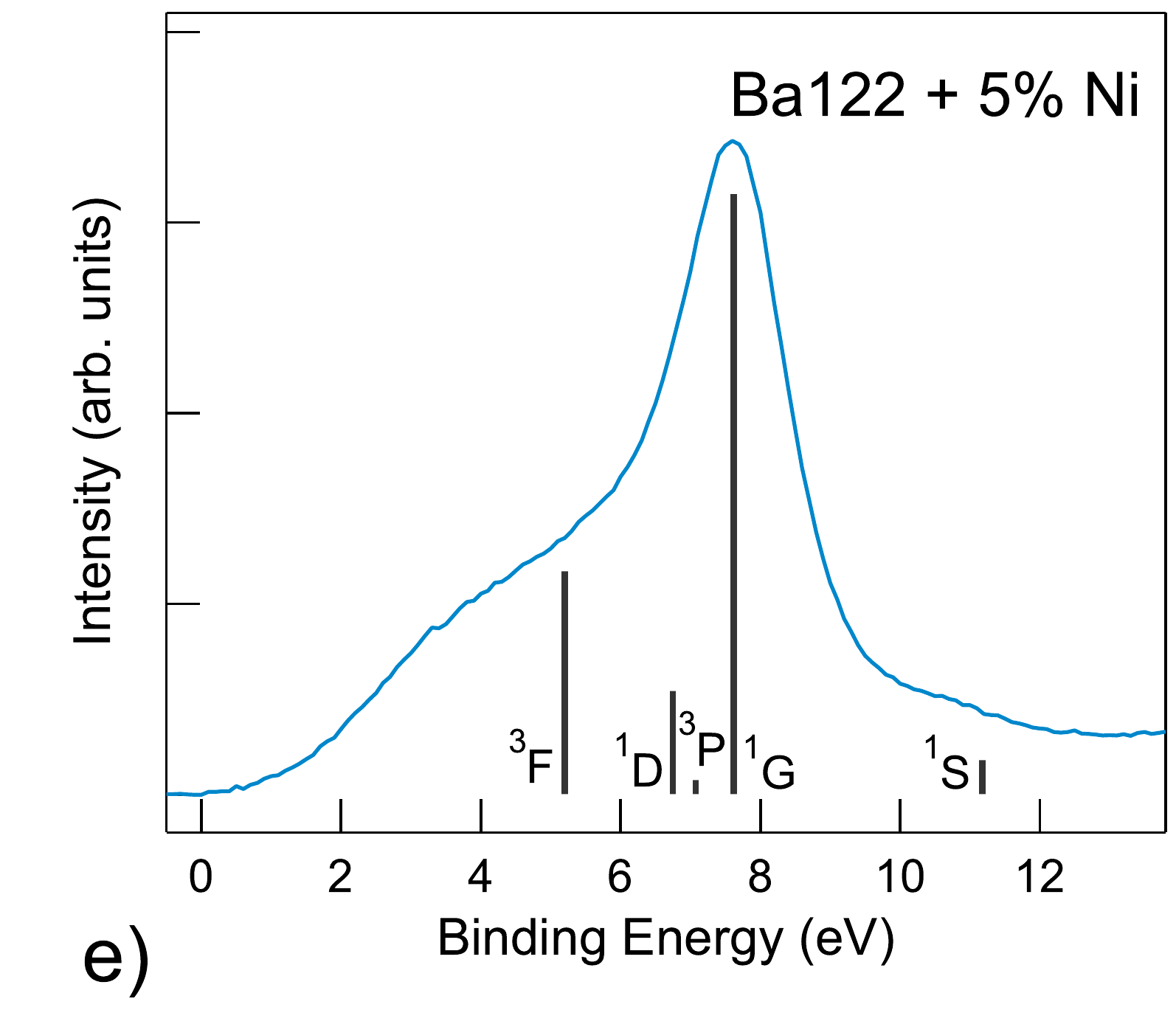}
\includegraphics[width=0.32\textwidth]{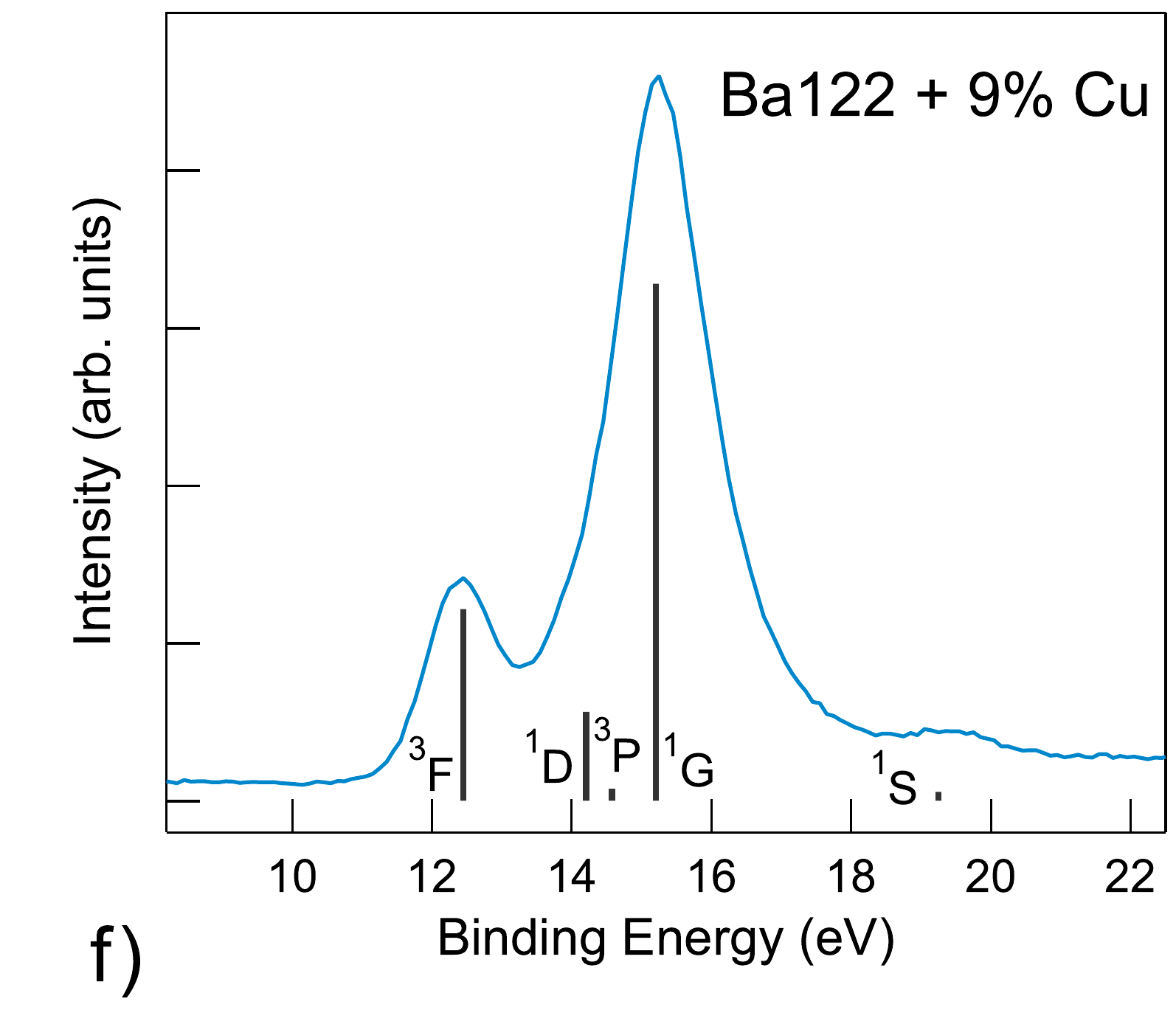}
\caption{Resonant PES scans across the L$_3$ edge for a) Ca(Fe,Co)$_2$As$_2$, b) Ba(Fe,Ni)$_2$As$_2$, and c) Ba(Fe,Cu)$_2$As$_2$. On the left side of each map in the top row the XAS spectra are displayed. From each map a off-resonance PES spectra was subtracted, which is shown below each map. The AES peak for each incident photon energy is indicated by black dots, showing the crossover from a resonant Raman ($\epsilon_B=$const.) to the normal AES decay ($\epsilon_K=$const.) . The lower panels show $LVV$ Auger spectra corresponding to the cuts indicated by the dashed lines in the upper panels. Term splittings of the atomic $3d^8$ final state were calculated using Cowan's code\,\cite{Cowan1981} and the AES intensities were taken from Refs.\,\onlinecite{Bennett1983} and \onlinecite{McGuire1977} (vertical lines). The calculations do not take into account the full resonance process, but consider the Auger decay of the core hole state only.}
\label{fig:Res_PES}
\end{figure*}

The PES data in Fig.\ref{fig:PES} is experimental proof that the $3d$ impurity states move to higher binding energy with increasing atomic number $Z$. While the Ni states around 1.4\,eV are still located inside the Fe bands, although close to the band bottom,
% of the Fe band at 2\,eV. 
the Cu impurity states are well below the Fe bands. The lack of additional features for the Co-substituted material can be attributed  to the fact that these states hybridize strongly with the Fe bands and are distributed over essentially the same energy region.% as the Fe states.

The above results are in good agreement with previous reports, where similar trends have also been extracted\,\cite{Wadati2010,McLeod2012}.
An important new observation here is that replacing Fe by heavier Ni and Cu, removes Fe-states close to the Fermi level and adds impurity states at higher binding energy, as is nicely demonstrated by the data in Fig.\,\ref{fig:PES}. The PES data thereby shows that the heavier TMs contribute less to the states close to the Fermi level. Note that these data can only provide qualitative information about changes in the density of states (DOS), because the PES cross sections for Fe, Co, Ni and Cu are different and not known precisely.

Resonant PES data for the three substituted samples are presented in Fig.\,\ref{fig:Res_PES}, including the XAS and the non-resonant PES. The intensity maps as a function of photon energy ($h\nu$) and binding energy ($\epsilon_B$) have been obtained by subtracting  the non-resonant PES spectrum, in order to highlight the resonant features. In all cases a strong increase of the PES intensity can be observed, as soon as the photon energy is tuned to the TM-impurity L$_3$ edge. The PES maps in Fig.\,\ref{fig:Res_PES} show the typical Auger features of resonant PES, namely (i) a resonant Raman Auger decay right at the TM L$_3$ threshold, which occurs at a fixed binding energy, and (ii) the conventional Auger decay at higher $h\nu$, which leads to a constant kinetic energy of the emitted Auger electron ($\epsilon_{kin}^{AES}$) and therefore shifts linearly when plotted as a function of $\epsilon_B=h\nu-\epsilon_{kin}^{AES}$\,\cite{Gelmukhanov1999,Levy2012}. Comparing the PES maps for Co, Ni and Cu, it can immediately be observed that the Auger final states move to higher binding energies with increasing $Z$ and, at the same time, develop a clear fine structure.

We will consider the following processes: $3d^n+h\nu \rightarrow \underline{2p}3d^n + e^*   \rightarrow 3d^{n-2} +k+ e^* $, with $\underline{2p}$ the $2p$ core hole, $e^*$ the corresponding photo-excited electron, and $k$ the emitted Auger electron. The first step corresponds to the absorption of the incoming photon and the second step is the $LVV$ Auger decay. Note that $e^*$ does not participate in the $LVV$ Auger decay, resulting in 2 additional holes in the $3d$ valence shell.
In the case of an isolated atom, the interaction between these holes splits the Auger final state into a characteristic multiplet, consisting of different terms with different energy, total spin $S$ and angular momentum $L$.  Each of these final $LS$-terms gives rise to a line in the total spectrum with a specific intensity. If these lines are sharp and well defined, the multiplet structure of the AES provides a unique fingerprint of the electronic final state configuration.

The broad AES spectrum of Co, however, does not show any clear multiplet structure, as can be observed in Figs.\,\ref{fig:Res_PES}\,a) and d). This implies that the Co-states are strongly hybridized with the Fe-host: the state with the 2 holes on Co couples strongly to the continuum of states where one hole has moved from Co into the Fe-host, resulting in the observed broad and structureless spectrum. This observation is in agreement with a recent resonant PES study on \caco\,\cite{Levy2012}.

As can be seen in Figs.\,\ref{fig:Res_PES}\,b) and e), Ni shows a much more structured $LVV$ spectrum. The observed lineshape is essentially identical  to that found for Ni-metal\,\cite{Weinelt1997,Bennett1983}, showing structures that agree very well with an atomic Ni:$3d^8$ Auger final state multiplet.  Both the energy splitting between the different terms and the relative intensities obtained by the atomic calculation\,\cite{Bennett1983} are in very good agreement with the experimental data. 

The  observation of a $3d^8$ Auger final state for Ni is remarkable, because it is incompatible with the often assumed electron doping. The latter would result in a Ni:$3d^6$ in the ground state, from which the $3d^8$ Auger final state cannot be reached at all. A Ni:$3d^6$ is hence excluded by our data.
Instead, the data  implies a strong Ni:$3d^9$ component in the ground state. Here the $3d^8$ final state can be reached via  $3d^9+h\nu\rightarrow\underline{2p}\,3d^{10}\mathrm{(screening)}+e^*\rightarrow3d^8+e^*+k$, where the $3d^{10}$ in the intermediate state is due to the screening of the core hole. In principle, a Ni:$3d^{10}$ in the ground state would be also possible. But we can exclude a strong  Ni:$3d^{10}$ based on the XAS data, where we and others\,\cite{McLeod2012} observe a strong XAS peak at the Ni $L_3$ edge, which would be absent for a $3d^{10}$ configuration of Ni. 

The $LVV$ spectrum for Cu is even more structured, as demonstrated in Figs.\,\ref{fig:Res_PES}\,c) and f).  Again, we compare the experimental results to an atomic multiplet calculation \cite{McGuire1977} and find excellent agreement with a Cu:$3d^8$ final state multiplet. This identification is further supported by comparing our data to  previously published results on Cu-metal and CuO$_2$, which also show the Cu:$3d^8$ final state multiplet\,\cite{Foehlisch2001,Tjeng1991}. 
%Furthermore, the good agreement with the atomic calculation implies that, in striking contrast to the Co-case, the Cu-impurities do not hybridize significantly with the host.

The Cu AES data therefore unambiguously identify a significant Cu:$3d^{10}$ or Cu:$3d^9$ component in the ground state. In fact, since XAS shown in  Fig.\,\ref{fig:Res_PES}\,c) agrees very well with what is expected for a Cu:$3d^{10}$\,\cite{Grioni1992}, we can exclude a significant population of Cu:$3d^9$. This result is also in accord with the high $\langle\epsilon_B^{Cu}\rangle$ of the Cu:$3d$ states determined by the PES data in Fig.\,\ref{fig:PES}.  A Cu:$3d^{10}$ state was also deduced in a recent x-ray spectroscopy study\,\cite{McLeod2012} and indirectly concluded from macroscopic measurements on BaCu$_2$As$_2$\,\cite{Anand2012}. We can therefore safely conclude that Cu is predominantly in a Cu:$3d^{10}$ configuration, i.e., the Cu-impurities have a closed shell. 

\begin{figure}
\centering
%\vspace{8cm}
\includegraphics[width=0.9\columnwidth]{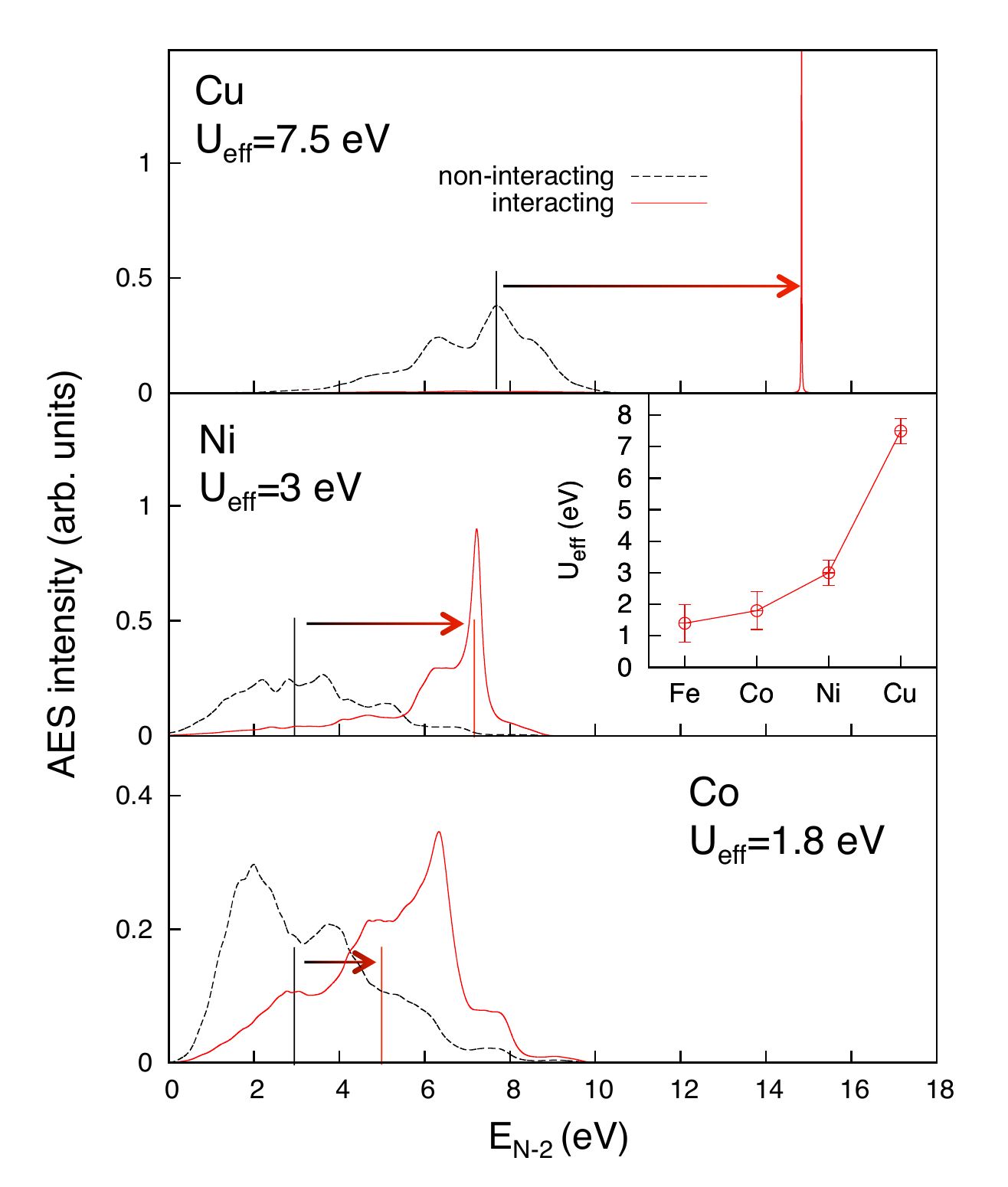}
\caption{Simulated AES spectra and estimated $U_{eff}$.  Non-interacting case:  The continuum described by $D(\epsilon)$ shifts to higher energies with increasing $Z$ due to the increased binding energy of the impurity states.  Interacting case: while no 2-hole bound state is formed for Co, the $U_{eff}$ is large enough to  push a 2-hole bond state  out of the $D(\epsilon)$-continuum.}\label{fig:CS}
\end{figure}
 
%% new paragraph
In the current situation, a realistic estimate can be made for the effective onsite Coulomb interaction $U_{eff}$ on the impurities, using the Cini-Sawatzky (CS) theory\,\cite{Sawatzky1977,Vos1984,Vilmercati2012}. Within this approach the AES intensity is %expressed as
%(see e.g. Ref.\,\cite{Vilmercati2012} and references therein) 
%
\begin{displaymath}
I(\epsilon)=\frac{D(\epsilon)}{\left(1-U_{eff}F(\epsilon)\right)^2+(\pi\,U_{eff}\,D(\epsilon))^2},
\end{displaymath}
where $\epsilon$ is the energy of the 2-hole Auger final state relative to the chemical potential, $D(\epsilon)=\int{\rho(x)\,\rho(\epsilon-x)dx}$, $F(\epsilon)=P\int{D(x)/(\epsilon-x)dx}$ and $\rho$ is the impurity DOS with the occupied part normalized to 1.  Here, we determined  $\rho$ by means of density functional theory. These band structure calculations were done in the local density approximation using the full potential code {\tt WIEN2K}\,\cite{Blaha2001}, as described in Ref.\,\onlinecite{Wadati2010}. The results of this calculation for non-interacting ($U_{eff}=0$, $I(\epsilon)=D(\epsilon)$)  and interacting ($U_{eff}\neq0$) holes are presented in Fig.\ref{fig:CS}.

For Fe (not shown) and Co we determined $U_{eff}$ by fitting the center of mass of the calculated spectra to the corresponding experimental values, which are $(4.0\pm 0.5)$\,eV for Fe and $(4.5\pm 0.5)$\,eV  for Co. For Ni and Cu, we set $U_{eff}$ as to match the measured binding energy of the $^1G$ term, which is 7.35\,eV and 14.85\,eV, respectively.

The energies $E_{N-2}$ of the Auger final states are expressed on the binding energy scale and were determined using the resonant Auger Raman spectra.
%The use of the well-calibrated binding energy scale minimizes experimental errors related to the beamline. However, in this case the excited electron $e^*$ is not emitted and can cause additional interactions that potentially affect the measured $E_{N-2}$.  But as we describe in the supplement, the effects due $e^*$ are found to be small.
%
In this way we obtain $U_{eff}=(1.4\pm0.6)$\,eV, $(1.8\pm0.6)$\,eV, $(3.0\pm0.4)$\,eV and $(7.5\pm0.4)$\,eV for Fe, Co, Ni and Cu (inset of Fig.\,\ref{fig:CS}). $U_{eff}$-values for Fe and Co were also reported previously\,\cite{Koitzsch2010,Vilmercati2012, Levy2012} and are in fair agreement with our results.
The analysis explains the development of the quasi-atomic $LVV$:
% in Fig.\,\ref{fig:Res_PES}: 
with increasing $U_{eff}$ the Auger final states are pushed out of the continuum of states given by $D(\epsilon)$ and form localized bound states\,\cite{Sawatzky1977}. The emerging multiplet structure with increasing $Z$ in Fig.\,\ref{fig:Res_PES} is therefore a direct consequence of the increasing $U_{eff}$. 

The $U_{eff}$ determined here corresponds to the effective Coulomb interaction between 2 holes in the valence shell.  $U_{eff}$ depends on the spatial structure of the 2-hole wave function and, most importantly, the screening of these holes by the surrounding charges. Especially the latter strongly reduces the bare (atomic) Coulomb repulsion of $\sim$25\,eV to the observed $U_{eff}$. The increase of $U_{eff}$ with $Z$  is consistent with the observed  charge accumulation, i.e., the filling of the TM:$3d$-shell with increasing $Z$, as this indeed reduces the number of possible screening channels (cf.\,Ref.\,\onlinecite{Leonov2006}). 
It is important to point out that the onsite repulsion parameter U used in various theories is model dependent, because different approximations for the wave functions are used and the screening is taken into account to different degrees. U can therefore differ from the fully screened $U_{eff}$ determined here. For example, U for a multi-band model, which takes into account all the relevant screening channels, would be close to the atomic value of $\sim$25\,eV.

We now discuss the consequences of the above results. Our data confirm the strong overlap and hybridization of the Co and Fe states. The Co states therefore have a significant Fe:3d character, i.e., Co adds electrons to the Fe bands and the number of charge carriers with Fe:3d character increases.
%, causing a shift  of the Fermi energy $\epsilon_F$ within the Fe bands.
%
But this does not mean that the average local electron density \nfe\/ at the Fe-sites increases accordingly, because the delocalized charge carriers have an increased probability of being around Co. Charge density therefore
% corresponding to these delocalized band states 
piles up around the Co\,\cite{Levy2012,Wadati2010,Berlijn2012}, resulting in \nfe$\simeq$const., as was indeed observed by local probes\,\cite{Bittar2011,Merz2012,Khasanov2011}. The scattering potential due to the Co-impurities is given by $\langle\epsilon_B^{Co}\rangle\simeq1.75\,\langle\epsilon_B^{Fe}\rangle$ ($\langle\epsilon_B^{Fe}\rangle=0.4$\,eV)\,\cite{Levy2012} and $U_{eff}^{Co}\simeq1.3\,U_{eff}^{Fe}$.

The Ni states are located close to the bottom of the Fe bands and the data in Fig.\,\ref{fig:PES} show that these impurity states contribute only little to the states close to $\epsilon_F$. % (the PES intensity between  0-1\,eV is reduced). 
At the same time, the  Ni:$3d^9$ %in the ground state 
observed here
is direct and clear experimental proof for the charge accumulation at Ni. As in the Co-case, this is perfectly in line with the \nfe$\simeq$const. observed by local probes\,\cite{Bittar2011}. 
%
%We also find that t
The scattering potential due to the Ni-impurities is much more severe than in the Co-case, because $\langle\epsilon_B^{Ni}\rangle\simeq3.5\,\langle\epsilon_B^{Fe}\rangle$ and $U_{eff}^{Ni}\simeq2.1\,U_{eff}^{Fe}$. This will affect the electronic structure and result in a  broadened FS.

The Cu impurities differ from the previous cases in that they form bound states well below the Fe bands with little Fe:3d character. 
%(cf. Fig.\,\ref{fig:PES}). 
%The extra electrons are therefore barely add to the Fe bands. 
The observed Cu:$3d^{10}$ configuration implies a strong charge accumulation at Cu, which may even result in hole doping of the Fe bands
% and a  reduction of \nfe\/
% which  Without any Fe:3d character of the impurity states \nfe\/ would actually be reduced by Cu-substitution 
(see also Ref.\,\cite{McLeod2012}).  
So the electron doping is reduced and possibly even turned into hole doping when going from Co over Ni to Cu.
We also find that the scattering due to Cu with its closed shell is very strong ($\langle\epsilon_B^{Cu}\rangle\simeq10\,\langle\epsilon_B^{Fe}\rangle$, $U_{eff}^{Cu}\simeq5.3\,U_{eff}^{Fe}$). Correspondingly, a few percent Cu substitution will already make the definition of a FS difficult. We note that the above is in line with a very recent neutron scattering study\,\cite{KimPRL12}. 
%Here the application of Luttinger's theorem is questionable.

%The number of additional electrons depends critically on the binding energy of the impurity states. 
To conclude, if and how many electrons the TM substitution adds to the Fe bands depends crucially on the impurity-host hybridization and the energy difference between the host and impurity states. These parameters depend on the chosen TM and their effects go beyond a mere charge carrier doping. Especially the additional scattering potentials, which are quantified here, affect the low-energy electronic structure, play a role for the phase diagrams and need to be considered in realistic theoretical models. % substituted Fe-pnictides and the discussion of their phase diagrams. 
% The latter two strongly depend on chosen TM. 
%Our results strongly support the notion that even if electrons with Fe:3d character are added, the charge accumulation at the TM impurities causes \nfe$\simeq$constant. HTS is therefore not primarily driven by changing \nfe. %Rather changes of the FS shape and broadening seem to be instrumental in creating HTS; probably via a suppression of the SDW. 
%Our results also quantify the strength of the random scattering potentials introduced by the TM substitution. 
Interestingly, the HTS upon Ni substitution proves that the HTS in the pnictides is robust against strong impurity scattering. 

Work at the IFW was supported by the DFG (Grant BE1749/13, SPP1458). SW, JG, and RK acknowledge the support by DFG through the Emmy-Noether program (Grant  WU595/3-1, GE1647/2-1). VB acknowledges financial support by the DAAD and GL, ISE and GAS from the Canadian funding agencies NSEC, SFI and CRC. We thank HZB for the allocation of synchrotron radiation beamtime and S. Krause, M. Oehzelt and S. Pohl for their support at the beamline.

%\bibliography{FeAs_paper}
%\bibliographystyle{apsrev4-1}
%merlin.mbs 2010-03-15 4.21a (PWD, AO, DPC)

\newpage

\section{Supplementary material:\\ Growth and characterization of Ba(Fe,TM)$_2$As$_2$ (TM=Co, Ni, Cu)}

\begin{figure}[h!]
\centering
%\vspace{15cm}
\includegraphics[width=0.85\columnwidth]{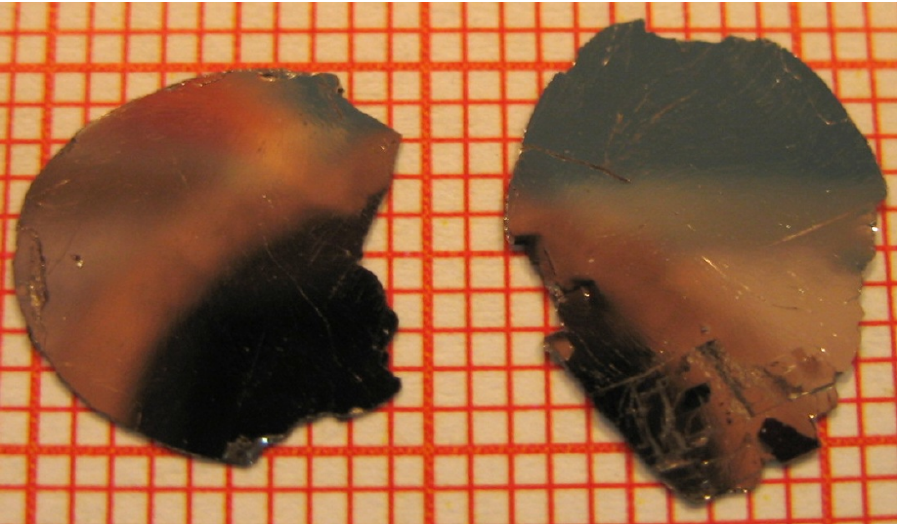}
\caption{As grown Ba(Fe$_{0.95}$Ni$_{0.05}$)$_2$As$_2$ single crystals.}\label{fig:crystals}
\end{figure}

All the crystals used for the present study were grown by the self flux technique as described in Refs.\,\onlinecite{Harnagea2011} and \cite{Aswartham2011a}, using FeAs as flux. In case of the Ba(Fe$_{1-x}$Co$_x$)$_2$As$_2$ system, details about the crystal growth, the characterization and the physical properties were published in Ref.\,\onlinecite{Aswartham2011a}. Here we provide additional details regarding the growth and the characterization of the Ni and Cu substituted BaFe$_2$As$_2$ single crystals. The pre-reacted precursor materials FeAs, Fe$_2$As, BaAs and metallic Ni or Cu (TM) were mixed, leading to a Ba(Fe$_{1-x}$TM$_x$)$_{3.1}$As$_{3.1}$ composition. This composition was used to achieve a homogeneous melt at T = 1463 K. The melt was cooled slowly under a temperature gradient in a double-wall crucible assembly to obtain large and flux-free single crystals of Ba(Fe$_{1-x}$Ni$_x$)$_2$As$_2$ and Ba(Fe$_{1-x}$Cu$_x$)$_2$As$_2$. Two examples for the obtained single crystals with typical dimensions are shown in the Fig\,\ref{fig:crystals}.

The high quality of the grown single crystals was assured by several complementary techniques. From each batch, several samples were examined with a Scanning Electron Microscope (SEM Philips XL 30) equipped with an electron microprobe analyzer for the semi-quantitative elemental analysis in the energy dispersive X-ray (EDX) mode. Using EDX, the Ni and Cu concentration was determined by averaging over several different points on the sample surface. The estimated composition from the EDX for the present single crystals is Ba(Fe$_{0.95}$Ni$_{0.05}$)$_2$As$_2$ and Ba(Fe$_{0.91}$Cu$_{0.09}$)$_2$As$_2$. In general, the error of an EDX analysis is about 2 mass percent without any additional standardization procedure. However, the method of averaging reduces the size of the error bars significantly. In this way the concentration $x$ of Co, Ni and Cu was determined to be $(0.06\pm0.01)$, $(0.05\pm0.01)$ and $(0.09\pm0.01)$. 

\begin{figure}[t!]
\centering
%\vspace{15cm}
\includegraphics[width=0.95\columnwidth]{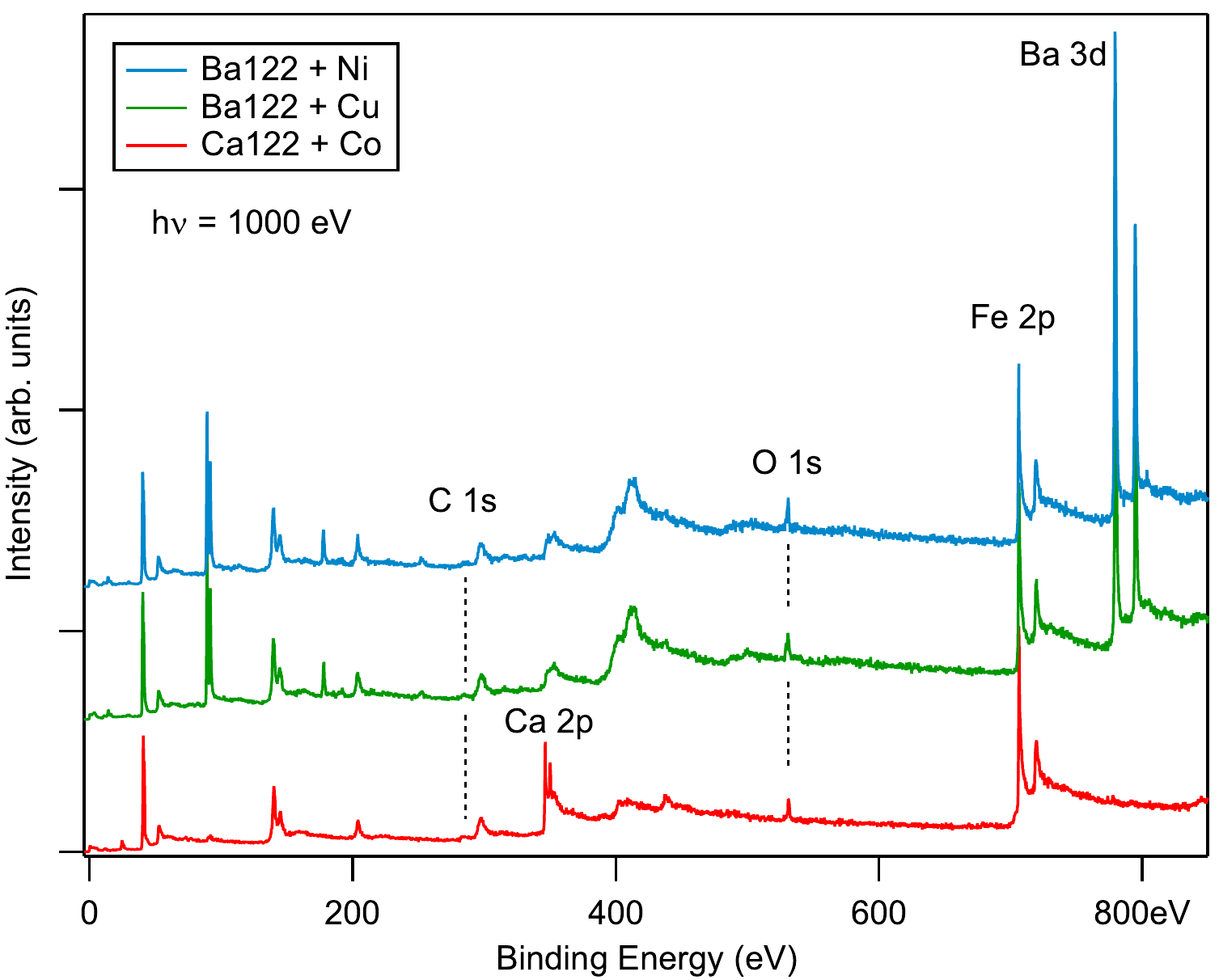}
\caption{XPS survey scans. Only a small trace of oxygen is observed, which can at least partially be attributed to adsorbates on the sample surface and verifies the high purity of our samples.}\label{fig:survey}
\end{figure}

In addition to this, the TM contents of the studied samples were also verified in-situ by XPS measurements, which gave the same results as EDX. The in-situ XPS measurements also do not show any sign of a significant oxygen contamination in any of the studied samples. Representative XPS survey scans are displayed in Fig.\,\ref{fig:survey}.  The weak O 1s line is comparable to what has been reported previously in the literature for high-purity samples\,\cite{McLeod2012}, which verifies the high quality of the studied single crystals. 
%The O traces can at least partially be attributed to oxygen containing adsorbates on the sample surface, which, however, have no influence on our data. 
Our XAS measurements also exclude the presence of Fe-oxide impurities, which would leave a clear fingerprint in these spectra\,\cite{Merz2012}.

\begin{figure}[b!]
\centering
%\vspace{15cm}
\includegraphics[width=0.85\columnwidth]{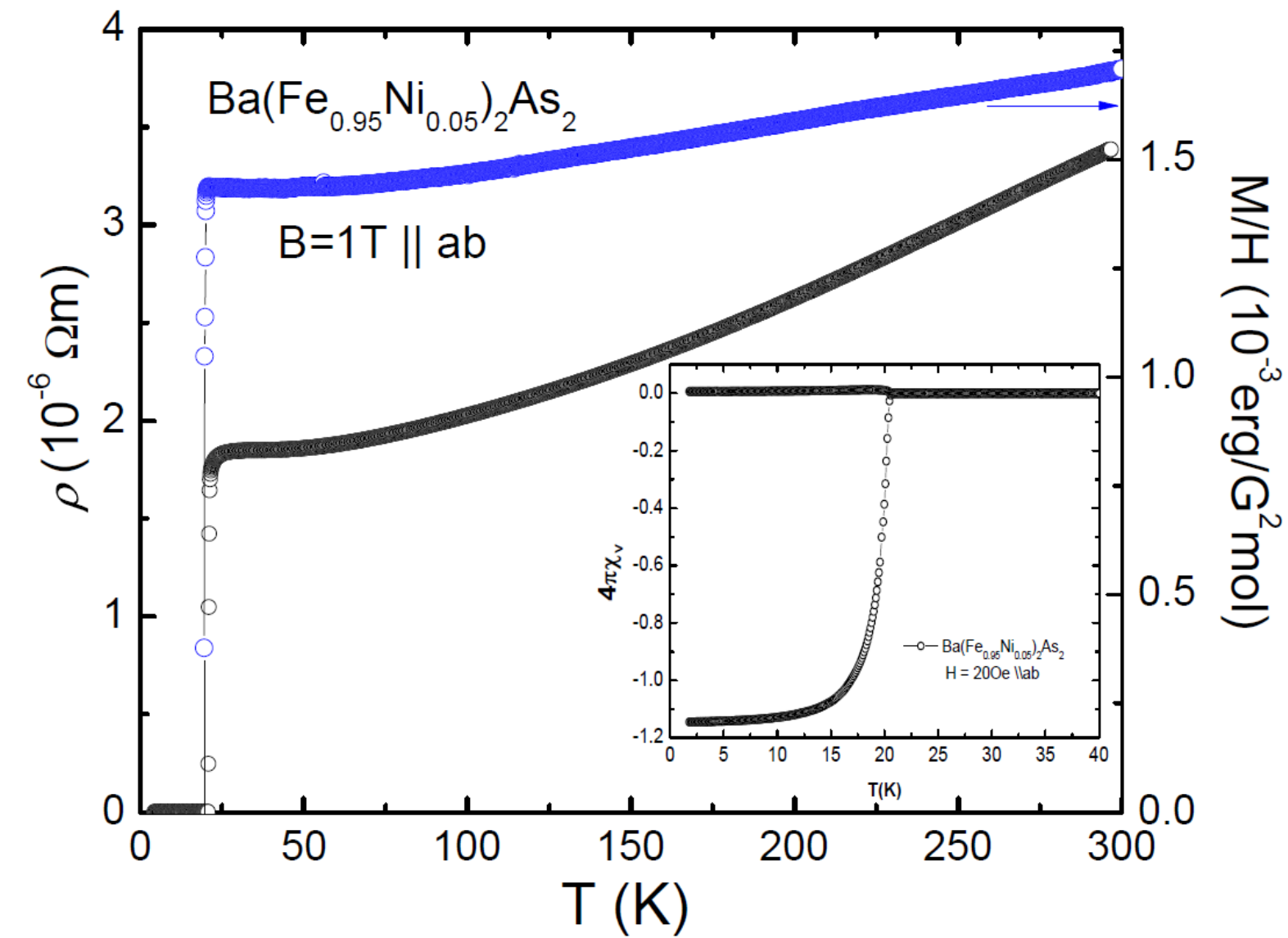}
\caption{Magnetization and resistivity measurement of Ba(Fe$_{0.95}$Ni$_{0.05}$)$_2$As$_2$ single crystals}
\label{fig:macos}
\end{figure}

Prior the synchrotron experiments, the high quality of our single crystals was also checked by XRD, resisitivity and mangnetization measurements. In Fig.\,\ref{fig:macos} we show representative measurements for our Ba(Fe$_{0.95}$Ni$_{0.05}$)$_2$As$_2$ single crystals. The data in Fig.\,\ref{fig:macos}, reveals that the structural and magnetic transitions are completely suppressed by 5\% Ni-substitution, whereas bulk superconductivity occurs. This is in excellent agreement with the literature (S.L. Bud'ko {\it et al.}, Phys. Rev. B {\bf 79}, 220516R (2009)). Furthermore, the superconducting transition at Tc = 21 K is very sharp and, within the error bars of our experiment, we find a superconducting volume fraction of 100\%. This further confirms the very high quality of our single crystals. 
Also the Cu-substituted samples were characterized by the same methods prior the synchrotron experiments. The structural quality was verified by XRD and, in agreement with the previous literature, the sample with 9\% Cu substitution does not show superconductivity down to 1.8 K.

\end{document}